# Calculation of Beam Propagation through a Defected or a Misaligned Two-Lens System

**Tariq Shamim Khwaja[1] and Syed Azer Reza[1]**

1. Department of Electrical Engineering, Lahore University of Management Sciences, DHA, Lahore 54792, Pakistan

## Abstract

An inadvertent or unwanted angular deviation to a passing beam can be introduced by any single optical component within an optical system. The problem arises due to an imperfect tilt alignment of the optical component or manufacturing defects which results in a slightly different response than expected from an ideally aligned/manufactured component. The resulting beam deviation can plague any well-designed optical system that assumes the use of ideal components. The problem is especially introduced by non-ideal lenses, transparent plates and optical windows. Here we present a simple method of manually introducing a deviation error angle which we assume to be constant for all beam incidence angles over the paraxial range.

## 1. Introduction

Tracing the path of a beam through optical elements generally requires the use of the ABCD matrices [1]. This method is used for propagation through optical elements which are symmetric about a common optical axis (generally referred to as the optical axis of the system). However, due to either manufacturing defects or imperfect alignment, this assumption is not entirely true because defected or misaligned optical elements are not symmetric about the optical axis.

In this paper we propose a basic model for incorporating known angular deviations of lenses into our existing framework of ABCD matrices that is used for analyzing and computing beam propagation through an optical system.

## 2. Incorporating Angular Deviations due to Imperfect Parallelism of Surfaces in Standard ABCD Computations

We take the example of Electronically Controlled Tunable Lenses (ECTLs) [2-4]. Some tunable lens designs generally consist of an outer casing of an elastic membrane with liquid trapped inside this membrane-based encapsulation. Through design or inadvertently due to manufacturing defects, the two transparent faces of a lens (i.e. the entrance and the exit apertures of a lens) can suffer from the problem of an imperfect parallelism. Such imperfect parallelism of surfaces, as shown in Fig.1, cause an extra angular ray deviation which is unaccounted for in standard ABCD models.

In this paper, we introduce an extra correction angle to the angular deviation that a beam would experience from an equivalent ideal optical component. The method involves simply introducing an additional error angle to a beam which would otherwise pass through an equivalent ideal optical component.

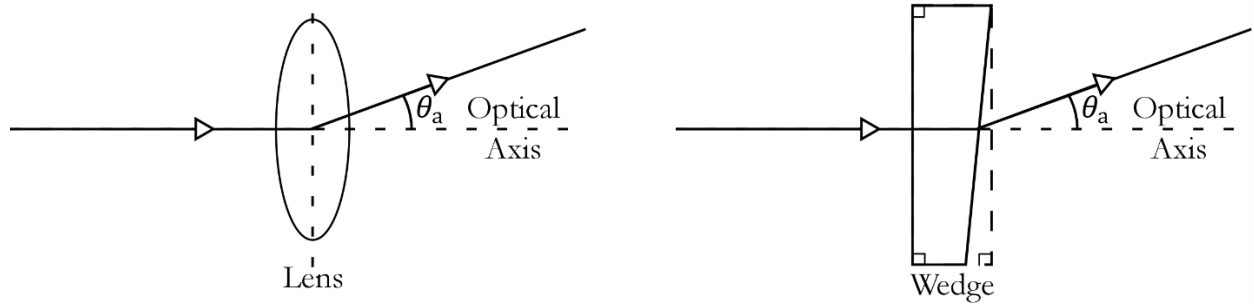

Fig.1 Beam angular deviation due to manufacturing defects and imperfect parallelism of surfaces

### 2.1. An Afocal System with one Defected or Misaligned Lens

Telescopic imaging systems [5-6], beam correction systems [7] and opthalmometers [8] often deploy an afocal system which involves the use of two lenses separated by a distance which is equal to the sum of their focal lengths. Due to the widespread use of afocal systems, as an example of a slightly defected optical system, we first consider a 2-lens afocal system in which first lens (Lens 1) introduces a fixed additional angular deviation irrespective of the height or incidence angle of an incoming ray.

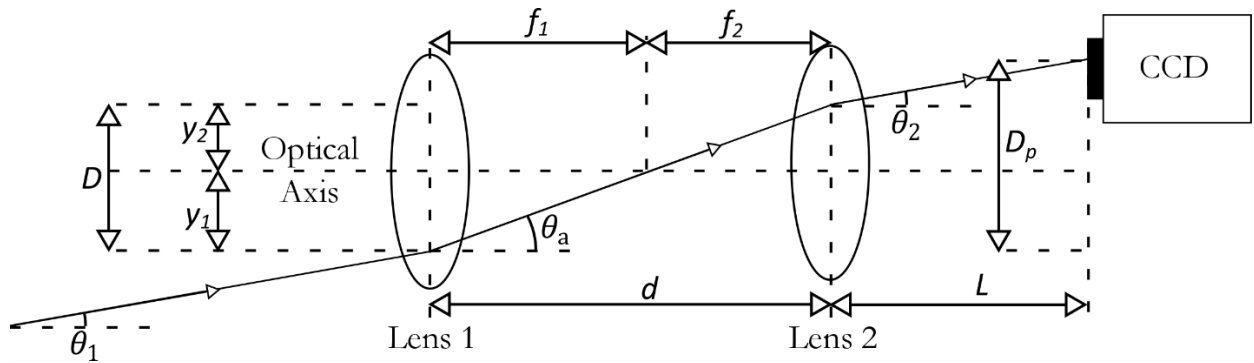

Fig.2 A Two-Lens System with a Defected First Lens

An incoming beam is incident on the system with an initial beam height $y_1$ and an incidence angle $\theta_1$, as shown in Fig.2. The ABCD matrix of Lens 1 with focal length $f_1$, is given by:

$$\begin{bmatrix} A & B \\ C & D \end{bmatrix}_{Lens1} = \begin{bmatrix} 1 & 0 \\ -1/f_1 & 1 \end{bmatrix} \tag{1}$$

The incident beam exits with a height $y_a$ and angle $\theta_a'$ after passing through Lens 1. Here $y_a$ and $\theta_a'$ of the emerging rays are calculated for a perfect lens which does not introduce any additional angular deviation due to an imperfect parallelism.

$$\begin{pmatrix} y_a \\ \theta_a' \end{pmatrix} = \begin{pmatrix} 1 & 0 \\ -1/f_1 & 1 \end{pmatrix} \begin{pmatrix} y_1 \\ \theta_1 \end{pmatrix} = \begin{pmatrix} y_1 \\ -1/f_1 \, y_1 + \theta_1 \end{pmatrix}$$

(2)

To account for an additional angular deviation $\theta_e$, we add an angle $\theta_e$ to $\theta_a'$ to determine the actual angle of the exit ray $\theta_a$.

$$\begin{pmatrix} y_a \\ \theta_a \end{pmatrix} = \begin{pmatrix} y_a \\ \theta_a' \end{pmatrix} + \begin{pmatrix} 0 \\ \theta_e \end{pmatrix} = \begin{pmatrix} y_1 \\ -1/f_1 \, y_1 + \theta_1 \end{pmatrix} + \begin{pmatrix} 0 \\ \theta_e \end{pmatrix} = \begin{pmatrix} y_1 \\ -1/f_1 \, y_1 + \theta_1 + \theta_e \end{pmatrix}$$

(3)

For an afocal system, i.e. two lenses of focal lengths $f_1$ and $f_2$ which are separated by a distance of $f_1 + f_2$, the ABCD matrix for beam propagation between the two lenses is given by:

$$\begin{pmatrix} 1 & f_1 + f_2 \\ 0 & 1 \end{pmatrix}$$

(4)

Beam propagation through Lens 2 with perfectly parallel interfaces is given by:

$$\begin{pmatrix} 1 & 0 \\ -1/f_2 & 1 \end{pmatrix}$$

(5)

Therefore, after propagating through Lens 1 and Lens 2, we can find the magnitude of the lateral shift $y_2$ and the exit angle $\theta_2$ of the beam with respect to the optical axis.

$$\begin{pmatrix} y_2 \\ \theta_2 \end{pmatrix} = \begin{pmatrix} 1 & 0 \\ -1/f_2 & 1 \end{pmatrix} \begin{pmatrix} 1 & f_1 + f_2 \\ 0 & 1 \end{pmatrix} \begin{pmatrix} y_a \\ \theta_a \end{pmatrix}$$

(6)

The overall ABCD matrix for the two-lens afocal system is given by:

$$\begin{pmatrix} y_2 \\ \theta_2 \end{pmatrix} = \begin{pmatrix} 1 & 0 \\ -1/f_2 & 1 \end{pmatrix} \begin{pmatrix} 1 & f_1 + f_2 \\ 0 & 1 \end{pmatrix} \begin{pmatrix} y_1 \\ -1/f_1 \, y_1 + \theta_1 + \theta_e \end{pmatrix}$$

(7)

Multiplying the three individual ABCD matrices (one for each lens and one for the separation between the lenses), we obtain:

$$\begin{pmatrix} y_2 \\ \theta_2 \end{pmatrix} = \begin{pmatrix} 1 & 0 \\ -1/f_2 & 1 \end{pmatrix} \begin{pmatrix} y_1 + (f_1 + f_2)\left(-1/f_1 \, y_1 + \theta_1 + \theta_e\right) \\ -1/f_1 \, y_1 + \theta_1 + \theta_e \end{pmatrix}$$

(8)

$$\begin{pmatrix} y_2 \\ \theta_2 \end{pmatrix} = \begin{pmatrix} 1 & 0 \\ -1/f_2 & 1 \end{pmatrix} \begin{pmatrix} (\theta_1 + \theta_e)f_1 + (\theta_1 + \theta_e)f_2 - f_2/f_1\, y_1 \\ -1/f_1\, y_1 + \theta_1 + \theta_e \end{pmatrix}$$

(9)

$$\begin{pmatrix} y_2 \\ \theta_2 \end{pmatrix} = \begin{pmatrix} (\theta_1 + \theta_e)f_1 + (\theta_1 + \theta_e)f_2 - f_2/f_1\, y_1 \\ (-1/f_2)\left((\theta_1 + \theta_e)f_1 + (\theta_1 + \theta_e)f_2 - f_2/f_1\, y_1\right) + (1)\left(-1/f_1\, y_1 + \theta_1 + \theta_e\right) \end{pmatrix}$$

(10)

$$\begin{pmatrix} y_2 \\ \theta_2 \end{pmatrix} = \begin{pmatrix} (\theta_1 + \theta_e)f_1 + (\theta_1 + \theta_e)f_2 - f_2/f_1\, y_1 \\ -f_1/f_2\,(\theta_1 + \theta_e) - (\theta_1 + \theta_e) + 1/f_1\, y_1 - 1/f_1\, y_1 + (\theta_1 + \theta_e) \end{pmatrix}$$

(11)

A simplified form for the height and exit angle of the beam is hence given by:

$$\begin{pmatrix} y_2 \\ \theta_2 \end{pmatrix} = \begin{pmatrix} (\theta_1 + \theta_e)f_1 + (\theta_1 + \theta_e)f_2 - f_2/f_1\, y_1 \\ -f_1/f_2\,(\theta_1 + \theta_e) \end{pmatrix}$$

(12)

For $\theta_1 = 0$ (i.e. incoming beam is parallel to the optical axis)

$$\Rightarrow \begin{pmatrix} y_2 \\ \theta_2 \end{pmatrix} = \begin{pmatrix} \theta_e f_1 + \theta_e f_2 - f_2/f_1\, y_1 \\ -f_1/f_2\, \theta_e \end{pmatrix}$$

(13)

A negative value of $y_2$ indicates beam crossing the optical axis before reaching Lens 2. It is also clear from Eq.13 that if Lens 1 does not suffer from imperfect parallelism, i.e. $\theta_e = 0$, then the beam would simply exit the afocal system parallel to the optical axis, as is expected from a two lens system with no issues of imperfect parallelism.

## 2.2. Lateral Displacement Due to Angular Beam Steering

The lens separation for an afocal system is stated in Eq.4 as:

$$d = f_1 + f_2 \tag{14}$$

$$\Rightarrow f_2 = d - f_1 \tag{15}$$

The lateral displacement or shift, $D$, of the beam after passing through the two-lens system is calculated as

$$D = y_1 - y_2 = y_1 - \theta_e f_1 - \theta_e f_2 + \frac{f_2}{f_1} y_1 = y_1 - \theta_e f_1 - \theta_e (d - f_1) + \frac{(d - f_1)}{f_1} y_1 = y_1 - \theta_e d + \frac{d}{f_1} y_1 - \frac{f_1}{f_1} y_1 \tag{16}$$

$$D = \frac{dy_1}{f_1} - \theta_e d \tag{17}$$

The resulting beam displacement $D_P$ at any given plane located at a distance $L$ after Lens 2 along the optical axis is given by:

$$D_P = D + L\tan\theta_2 = \frac{dy_1}{f_1} - \theta_e d + L\tan\left(-\frac{f_1}{f_2}\theta_e\right) \tag{18}$$

$$\Rightarrow D_P = \frac{dy_1}{f_1} - \theta_e d + L\tan\left(\frac{f_1 \theta_e}{(f_1 - d)}\right) \tag{19}$$

## 2.3. An Afocal System with two Defected Lenses

It is evident from Eq.12 that the fixed angular deviation in the paraxial approximation simply adds to the incidence angle of the incoming beam with respect to the optical axis. If the second lens also imparts an additional angular deviation $\theta_{e2}$, then the height and angle of the beam exiting the afocal two-lens system is given by:

$$\begin{pmatrix} y_2 \\ \theta_2 \end{pmatrix} = \begin{pmatrix} (\theta_1 + \theta_e + \theta_{e2})f_1 + (\theta_1 + \theta_e + \theta_{e2})f_2 - \frac{f_2}{f_1} y_1 \\ -\frac{f_1}{f_2}(\theta_1 + \theta_e + \theta_{e2}) \end{pmatrix}$$

$$\tag{20}$$

For an incoming ray parallel to the optical axis i.e. $\theta_1 = 0$, Eq.20 simplifies to:

$$\begin{pmatrix} y_2 \\ \theta_2 \end{pmatrix} = \begin{pmatrix} (\theta_e + \theta_{e2})f_1 + (\theta_e + \theta_{e2})f_2 - \frac{f_2}{f_1} y_1 \\ -\frac{f_1}{f_2}(\theta_e + \theta_{e2}) \end{pmatrix}$$

$$\tag{21}$$

## 3. Conclusion

We have developed a theoretical framework to compute beam propagation through lenses with parallelism defects. We modified an existing ABCD matrix for an ideal lens and incorporated non-parallel surfaces through the introduction of an angular deviation term due to non-parallel interfaces. This angular deviation adds linearly to the deviation that an ideal lens would otherwise introduce to an incoming beam. In doing so, the effects of imperfect parallelism have been accounted for in the modified beam propagation model that we propose.